\documentclass[11pt]{amsart}
\usepackage{amscd,amssymb}
\usepackage{amsmath}
\usepackage{amsthm}

\usepackage[scaled]{helvet} 
\usepackage[pdftex]{graphicx,color}

\usepackage[cmtip,matrix,arrow]{xy}
\numberwithin{equation}{section}

\newtheorem {thm} 			{Theorem}     

\newtheorem {Lemma}     [equation]     	{Lemma}

\newtheorem {Conjecture}[equation]	{Conjecture}
\newtheorem {prop}      [equation]      {Proposition}

\theoremstyle{definition}
\newtheorem {defn}      [equation]      {Definition}
\newtheorem {rem}       [equation]	{Remark}

\newcommand{\pr} {\smallskip\noindent{\bf Proof\,\,}}

\DeclareMathOperator{\Z}{\mathbb Z}
\newcommand{\C}{\mathbb {C}}
\newcommand{\R}{\mathbb {R}}

\newcommand{\fg}{\mathfrak{g}}

\newcommand{\cA} {\mathcal{A}}
\newcommand{\cD} {\mathcal{D}}
\newcommand{\cF} {\mathcal{F}}

\begin{document}

\title[Fermionization and the Yang-Mills Quantum Field Theory]{Fermionization, Convergent Perturbation Theory, and Correlations in the Yang-Mills Quantum Field Theory in four dimensions}

\author{Jonathan Weitsman}
\address{Department of Mathematics, Northeastern University, Boston, MA 02115}
\curraddr{}
\email{j.weitsman@neu.edu}
\thanks{Supported in part by NSF grant DMS 04/05670}
\thanks\today


\date{}

\begin{abstract}
We show that the Yang-Mills quantum field theory with momentum and spacetime cutoffs in four Euclidean dimensions is equivalent, term by term in an appropriately resummed perturbation theory, to a Fermionic theory with nonlocal interaction terms.  When a further momentum cutoff is imposed, this Fermionic theory has a convergent perturbation expansion.  To zeroth order in this perturbation expansion, the correlation function $E(x,y)$ of generic components of pairs of connections is given by an explicit, finite-dimensional integral formula, which we conjecture will behave as 
$$E(x,y) \sim |x - y|^{-2 - 2 d_G},$$
\noindent for $|x-y|>>0,$ where $d_G$ is a positive integer depending on the gauge group $G.$  In the case where $G=SU(n),$ we conjecture that
$$d_G = {\rm dim~}SU(n) - {\rm dim~}S(U(n-1) \times U(1)),$$
\noindent so that the rate of decay of correlations increases as $n \to \infty.$

\end{abstract}

\maketitle
\section{Introduction}\label{sec:introduction}
Mathematical approaches to the path integrals of quantum field theory have in most cases not relied on perturbation theory, because of the expectation that the perturbation series, though well-defined, would diverge.  An exception has been the case of certain Fermionic theories \cite{fmrs,gk} (the Gross-Neveu model), where with spacetime and momentum cut-offs, perturbation theory is convergent,\footnote{In these models renormalized perturbation theory is {\em not} expected to be convergent.} and where the cut-offs can then be removed by renormalization group and cluster expansion methods.  In this paper we show that Yang-Mills theory in four Euclidean dimensions with spacetime and momentum cut-offs is equivalent to a Fermionic theory with a nonlocal polynomial interaction term, which, when further momentum cut-offs are placed on the Fermion fields, possesses a convergent perturbation expansion. We thus hope that the methods of \cite{fmrs,gk} will apply to allow removal of the momentum cut-off.  We show one might expect that the correlation functions in this theory have rapid polynomial decay to zeroth order in perturbation theory, which we hope will allow the infinite volume limit to be taken.  We also expect that if our cut-off theory correctly describes the behavior of the Yang-Mills quantum field theory, and if, as we conjecture, correlations in this Euclidean theory exhibit rapid polynomial decay, but not exponential decay, there should be consequences for lattice gauge theory and perhaps for experiment.

Fermionization is of course a familiar technique in two-dimensional quantum field theory \cite{c1,froh}, as well as in related areas of representation theory
\cite{fr}.  Its appearance here in the context of gauge theory is perhaps an indication of the ways in which path integrals contain surprises not apparent from their finite-dimensional analogs.

\subsection{The model}\label{themodel}
We begin with the classical Euclidean action for gauge theory in four dimensions.  Let $G$ be a compact simple Lie group and let $P$ be the trivialized principal $G$-bundle over $\R^4.$  Let $\cA(P)$ be the space of smooth, compactly supported connections on $P.$  We denote the coordinates on $\R^4$ by $(x_0,x_i),$ $i=1, 2 ,3,$ and identify $\cA(P)$ with $\fg$-valued one-forms on $\R^4.$  Choose an invariant metric on the Lie algebra $\fg.$  The Yang-Mills action $S: \cA(P) \to \R$ is given by 

$$ S(A) = \frac{1}{\lambda^2}\int_{\R^4}|F^+(A)|^2,$$
where $\lambda \in \R$ is the coupling constant and $F^+(A)$ is the self-dual part of the curvature of the connection $A.$\footnote{On connections with compact support on the trivial bundle, this action is equal to the more common form of the action $ S(A) = \frac{1}{\lambda^2}\int_{\R^4}|F(A)|^2.$} Since $S$ is invariant under the action of the gauge group, we must choose a slice for the action of the gauge group.  To do so we impose the axial gauge condition $A^0=0;$\footnote{
The advantage of axial gauge is that the Faddeev-Popov ghost terms decouple, so 
that the naive classical action gives rise to the correct quantum theory; see \cite{coleman}.  The absence of the (massless) ghost terms makes the long-range behavior of correlations more transparent.} to fix the remaining
time-independent gauge transformations,
 we impose the following conditions on $A$:\footnote{We have not fixed the constant gauge transformations; however, since this remaining symmetry group is compact, it will not affect convergence of the terms in the perturbation expansion.}
\begin{align}  {A}^1(0,x_1,x_2,x_3)& = 0 {\rm ~ for~ all~}x_1,x_2,x_3 \in \R.\notag\\{A}^2(0,0,x_2,x_3) &=0 {\rm~ for~ all~}x_2,x_3 \in \R.\label{gaugefix}
\\{A}^3(0,0,0,x_3)&=0{\rm ~for~ all~} x_3 \in \R. \notag \end{align}

We now impose a spacetime cutoff.  Let $\Lambda_0,\Lambda_1,\Lambda_2,\Lambda_3 > 0$ and
let 
$T^4=[-\Lambda_0,\Lambda_0]\times [-\Lambda_1,\Lambda_1]\times [-\Lambda_2,\Lambda_2]\times [-\Lambda_3,\Lambda_3].$
We restrict our attention to periodic connections, in other words, to connections living in the periodic box $T^4,$ and still obeying the gauge conditions $A^0=0$ and (\ref{gaugefix}).  Denote by $\cA$ the space of smooth $\fg$-valued one-forms $A$ on $T^4$ satisfying $A^0=0$ and the gauge conditions of equation
(\ref{gaugefix}).\footnote{It is not in general possible to impose the axial gauge conditions on connections on $T^4;$ for example, the winding number of the connection in the $0-$direction is an obstruction to such gauge fixing.  We do not expect this type of difficulty to make any difference in the infinite volume limit of a theory exhibiting clustering, as we shall see Yang-Mills may be expected to be. See also Remark \ref{instanton}.}  

To make sense of the perturbative quantum theory corresponding to this action, we introduce a momentum cutoff.  Let $\kappa > 0$ lie in the complement of the lattices $\pi \Z/\Lambda_i$, and let $\chi_{i,\kappa} \in C^\infty([-\Lambda_i,\Lambda_i])$ be the periodic function whose Fourier series is the function $\hat{\chi}_{i,\kappa}(\cdot)$ given by 
\begin{align*}
  \hat{\chi}_{i,\kappa}(x) = 1  {\rm~ if~}|x| < \kappa\\
 \hat{\chi}_{i,\kappa}(x) = 0  {\rm ~if~}|x| > \kappa. 
\end{align*}

Let $\chi_\kappa(x_0,x_1,x_2,x_3) := \prod_{i=0}^3 \chi_{i,\kappa}(x_i).$
Given $A \in \cA$ and $\kappa > 0$ in the complement of the lattices as above, let $A_\kappa= A \star \chi_\kappa,$ where $\star$ denotes convolution.
After a rescaling, our gauge fixed, cut-off action takes the form\footnote{In this paper we denote by $\nabla \times $ the three-dimensional projection
$\nabla \times = (\partial/\partial x_1) dx_1 + (\partial/\partial x_2) dx_2 + (\partial/\partial x_3) dx_3$ of the de Rham operator on one-forms on $T^4,$ and by $\nabla$ the three-dimensional projection of the de Rham operator on functions.  This notation agrees with calculus notation if we consider the gauge field $A$ and the conjugate field $F$ as $\fg$-valued vector fields on $T^4$  orthogonal to the $x_0$ direction.}

\begin{equation}\label{yma}
S_\kappa(A) = \int_{T^4} |d_0 A + \nabla \times A + \lambda [A_\kappa,A_\kappa]|^2.\end{equation}
Standard arguments then show that the Feynman perturbation series of the action
$S_\kappa$ consists of finite terms.  In the language of the physics literature, we have regularized the theory.

We may introduce a conjugate momentum $F \in \cF :=\Omega_{2}^+(T^4,\fg)$\footnote{We can consider $F$ as a vector on $T^4$ orthogonal to the $x_0$ direction
by taking the $i-$th component of the vector corresponding to $F$ to be $F^{0,i} + \sum_{j,k}\epsilon_{ijk}F^{j,k},$ $i=1,2,3.$} and consider the action 
$S_\kappa:\cA\times \cF \to \C$ given by $S_{H,\kappa}(A,F) = S_0(A,F) + S_{I,\kappa}(A,F),$ where\footnote{Here $\langle\cdot,\cdot\rangle$ denotes the $L_2$-inner product on $\fg \otimes \R^3$-valued functions arising from the chosen inner product on $\fg$ and the standard inner product on $\R^3,$ and $||\cdot||$ denotes the corresponding norm.}

$$S_0(A,F) = ||F||^2  -2 i \langle F, d_0A\rangle -2 i\langle F,\nabla \times A\rangle$$

$$S_{I,\kappa}(A,F) =-2 i \lambda\langle F, [A_\kappa, A_\kappa]\rangle.$$

Note that if we write $F_\kappa := F \star \chi_\kappa,$ then $S_{I,\kappa}(A,F) = -2i \lambda\langle F_{2 \kappa}, [A_\kappa, A_\kappa]\rangle.$

Let
$\cA^\kappa = \{A \in \cA:  A_\kappa = A\};$ similarly let $\cF^\kappa = \
\{F \in \cF:  F_{2\kappa} = F\}.$  Elements of 
$\cA^\kappa$ and $\cF^\kappa$ consist of forms with finite Fourier series.

We will check the following result in Section \ref{section:propagator}:
\begin{prop}\label{propagator1}
The quadratic form given by $S_0$ is nondegenerate on $(\cA^\kappa \times \cF^\kappa)\times (\cA^\kappa \times \cF^\kappa).$
\end{prop}
Since the action is quadratic in $F,$ standard arguments show that the actions $S_\kappa$ and $S_{H,\kappa}$ give rise to identical perturbation series.

Let $\Phi \in \fg \otimes \R^3$ denote the constant term of the Fourier series
of the conjugate field $F$; we will see that it plays a special role in the theory.  Let $\cF_0 = \{F \in \cF:  \int_{T^4} F = 0\}$ be the set of elements of $\cF$ with vanishing constant terms in their Fourier series. For any $\mu \in \R,$ let $S_{H,\kappa}^{\lambda, \mu}: \cA \times \cF_0 \times (\fg \otimes \R^3) \to \C$ be given by $S_{H,\kappa}^{\lambda, \mu}(A,F;\Phi) = S_{0,\kappa}^{\mu}(A,F;\Phi) + S_{I,\kappa}^{\lambda}(A,F),$

\noindent where 

$$S_{0,\kappa}^\mu(A,F;\Phi) = ||F||^2 -2 i\langle F, d_0 A\rangle -2 i \langle F, \nabla \times A\rangle -2 i \mu\langle\Phi,[A_\kappa, A_\kappa]\rangle,$$

\noindent and 

$$S_{I,\kappa}^{\lambda}(A,F) = -2 i \lambda \langle F, [A_{\kappa}, A_{\kappa}]\rangle.$$

Clearly if $\lambda = \mu$ the action $S_{H,\kappa}^{\lambda, \mu}$ coincides with the action $S_{H,\kappa}.$

Let $\cF_0^\kappa = \cF_0\cap \cF^\kappa.$  
As before we have the following proposition.

\begin{prop}\label{propagator}
Fix $\Phi \in \fg \otimes \R^3.$  The quadratic form given by $S_{0,\kappa}^\mu(\cdot,\cdot;\Phi)$ is nondegenerate on $(\cA^\kappa \times \cF_0^\kappa)\times (\cA^\kappa \times \cF_0^\kappa).$  The inverse of the corresponding matrix is bounded uniformly in $\Phi$ and uniformly in $\mu$ for $\mu \in \R.$
\end{prop}

Let ${C}_{\mu,\kappa}^{A,A}(\Phi)(x,y),$ $ {C}_{\mu,\kappa}^{A,F}(\Phi)(x,y),$ and $ {C}_{\mu,\kappa}^{F,F}(\Phi)(x,y)$ denote the components of the inverse of the matrix corresponding to the quadratic form $S_{0,\kappa}^\mu.$  It follows from Proposition \ref{propagator}  that these propagators are smooth functions on $T^4 \times T^4.$

To quantize this cut-off theory, we would like to make sense of the formal cut-off path integral

$$Z_\kappa(\lambda, \mu) := \int_{\fg \otimes \R^3} e^{-\frac12|\Phi|^2} d\Phi \int_{\cA^\kappa \times \cF^\kappa_0} \cD  A \cD  F e^{-\frac12 S_{H,\kappa}^{\lambda,\mu}(A,F;\Phi)}.$$

The integration over $\Phi$ is a finite dimensional Gaussian integral, independent of $\kappa,$ so we focus on the partition function
\begin{equation}\label{pf}
Z_\kappa(\lambda, \mu;\Phi) := \int _{\cA^\kappa \times \cF^\kappa_0} \cD  A \cD  F e^{- \frac12 S_{H,\kappa}^{\lambda,\mu}(A,F;\Phi)}.\end{equation}

Due to the ultraviolet and spacetime cutoffs, this path integral gives rise to a formal perturbation series in $\lambda,$ each of whose terms is finite.  We will show that with an additional ultraviolet cutoff, this perturbation series is {\em convergent} for all $\lambda \in \C$ for all $\mu \in \R.$

\begin{rem}
Note that the perturbation expansion for the partition function $Z_\kappa(\lambda,\lambda;\Phi)$ is not equivalent to the perturbation expansion of $Z_\kappa(\lambda)$; we have resummed an infinite number of terms.  In more physical language, the resummed perturbation expansion may be said to correspond to a different phase of the theory; the other phase most likely cannot be fermionized (see Remark \ref{ktoinf}).  From now on we will take the integral 

$$\int_{\fg \otimes \R^3} d\Phi e^{-\frac12 |\Phi|^2} Z_\kappa(\lambda,\lambda;\Phi)$$

\noindent as the {\em definition} of the partition function of the cut-off Yang-Mills theory.
\end{rem}

Choose an orthonormal basis $e_\alpha$ for $\fg,$ and let $f_{\alpha\beta\gamma}$ be the corresponding structure constants.  Let $({C}_{\mu,\kappa}^{A,A}(\Phi)(x,y))_{i,j;\alpha,\beta}$
denote the matrix elements of ${C}_{\mu,\kappa}^{A,A}(\Phi)(x,y)$ in the standard
basis for $\R^3$ and the chosen basis for $\fg;$ similarly we use the notation $({C}_{\mu,\kappa}^{A,F}(\Phi)(x,y))_{i,j;\alpha,\beta}$ and  $({C}_{\mu,\kappa}^{F,F}(\Phi)(x,y))_{i,j;\alpha,\beta}.$
 Let $\epsilon_{ijk}$ denote the antisymmetric three-tensor on $\R^3.$
Standard arguments give the following expression for the formal perturbation expansion of the path integral (\ref{pf}).  In this expression each of the terms is well-defined, but the power series may or may not be convergent.

\begin{defn}\label{pfdef}
The partition function $Z_\kappa(\lambda, \mu;\Phi)$ is given as a formal power series by
\begin{multline}\label{pfdefeqn}
Z_\kappa(\lambda, \mu;\Phi) := 
\exp\Bigl  (\sum_{i,j,\alpha,\beta} \int_{T^4\times T^4} dx dy 
 \Bigl [
({C}_{\mu,\kappa}^{A,A}(\Phi)(x,y))_{i,j;\alpha,\beta} 
\frac{\delta}{\delta J^i_\alpha(x)}\frac{\delta}{\delta J^j_\beta(y)}\\+ 2({C}_{\mu,\kappa}^{A,F}(\Phi)(x,y))_{i,j;\alpha,\beta} 
\frac{\delta}{\delta J^i_\alpha(x)}\frac{\delta}{\delta K^j_\beta(y)}\\+
({C}_{\mu,\kappa}^{F,F}(\Phi)(x,y))_{i,j;\alpha,\beta} 
\frac{\delta}{\delta K^i_\alpha(x)}\frac{\delta}{\delta K^j_\beta(y)}
 \Bigr ] 
\Bigr )\\
\exp \left( i \lambda \sum_{i,j,k; \alpha,\beta,\gamma}
\epsilon_{ijk}
f_{\alpha\beta\gamma}\int_{T^4} dx J^i_\alpha(x) J^j_\beta(x) K^k_\gamma(x)
\right)_{J=K=0}.
\end{multline}
\end{defn}

(Here the functions $J_\alpha^i(x),$ $K_\alpha^i(x),$ for $i = 1, 2, 3,$ $\alpha = 1, \dots, {\rm dim~}\fg,$ $x\in T^4,$ are formal variables.)

\subsection{Fermionization}

We now show that the partition function $Z_\kappa(\lambda, \mu;\Phi)$ is also the partition function of a Fermionic theory with nonlocal polynomial interaction.  Recall we have chosen a basis $e_\alpha$ for $\fg;$ for $\alpha = 1,\dots,{\rm dim~}\fg,$ let $\Psi_\alpha(x), \psi_\alpha(x)$ be complex Fermi fields on $T^4.$  Similarly, for $i=1,2,3,$ let $H_i(x) , \eta_i(x)$ be complex Fermi fields on $T^4.$  Consider the free Fermionic action

$$S_{F,0} (H_i,\Psi_\alpha, \eta_i,{\psi}_\alpha,\bar{H}_i,\bar{\Psi}_\alpha, \bar{\eta}_i,\bar{\psi}_\alpha)
:=
\int_{T^4} dx \sum_i |H_i(x)|^2 + \sum_i |\eta_i(x)|^2 
+\sum_\alpha |\Psi_\alpha(x)|^2 + \sum_\alpha |\psi_\alpha(x)|^2$$

\noindent and the interaction term
\begin{multline}\label{fermiint}
S^{\lambda,\mu}_{F,I,\kappa} (H_i,\Psi_\alpha, \eta_i,{\psi}_\alpha,\bar{H}_i,\bar{\Psi}_\alpha, \bar{\eta}_i,\bar{\psi}_\alpha)=\\
\int_{T^4 \times T^4} dx dy \sum_{\alpha,\beta,i,j}
 \Bigl [
H_i(x) \Psi_\alpha(x) ({C}_{\mu,\kappa}^{A,A}(\Phi)(x,y))_{\alpha,\beta; i,j} H_j(y) \Psi_\beta(y) +\\
2 H_i(x) \Psi_\alpha(x) ({C}_{\mu,\kappa}^{A,F}(\Phi)(x,y))_{\alpha,\beta; i,j\
} \eta_j(y) \psi_\beta(y) +\\
\eta_i(x) \psi_\alpha(x) ({C}_{\mu,\kappa}^{F,F}(\Phi)(x,y))_{\alpha,\beta; i,j} \eta_j(y) \psi_\beta(y)   \Bigr ]\\
+
(-i/2) \lambda \int_{T^4} dx \sum_{\alpha,\beta,\gamma;i,j,k} \epsilon_{ijk} f_{\alpha\beta\gamma} \bar{H}_i(x) \bar{\Psi}_\alpha(x)  
\bar{H}_j(x) \bar{\Psi}_\beta(x) 
\bar{\eta}_k(x)\bar{\psi}_\gamma(x).
\end{multline}

To make further progress, we impose a cutoff on the Fermi fields.  It is convenient to do this by convolutions with approximate delta functions and step functions, as follows.

Let $\zeta \in C^\infty(\R)$ be an even function satisfying

\begin{itemize}
\item $\zeta \geq 0.$
\item $\int_{-\infty}^\infty \zeta(x) dx  = 1$
\item $\zeta'(x) \leq 0$ for $x > 0.$
\item $\zeta(x) = 0$ for $x \geq 1.$
\end{itemize}

Given $\epsilon > 0,$ define $\delta_\epsilon: R^4  \to \R$ by

$$\delta_\epsilon(x_0,x_1,x_2,x_3) = (\frac{2}{\epsilon})^4\zeta(2x_0/\epsilon)\zeta(2x_1/\epsilon)\zeta(2x_2/\epsilon)\zeta(2x_3/\epsilon).$$

Similarly let $Z \in C^\infty(\R)$ satisfy

\begin{itemize}
\item $Z \geq 0.$
\item $Z(x) = Z(-x)$ for all $x \in \R.$
\item $Z(x) = 1$ if $x \in [-1,1]$
\item $Z'(x) \leq 0$ if $x > 0.$
\item $Z(x) = 0$ for $|x| \geq 2.$
\end{itemize}

For all $\epsilon > 0,$ define $D_\epsilon: \R^4 \to \R$ by

$$D_\epsilon(x_0,x_1,x_2,x_3)= Z(x_0/2\epsilon)Z(x_1/2\epsilon)Z(x_2/2\epsilon)Z(x_3/2\epsilon).$$

By periodizing the functions $\delta_\epsilon$ and $D_\epsilon,$ we obtain functions on the torus $T^4,$ which we continue to denote by $\delta_\epsilon$ and
$D_\epsilon.$

Define the cut-off Fermi fields by
\begin{align*}
\Psi_{\alpha,\epsilon} = \Psi_\alpha \star \delta_\epsilon\\
\bar{\Psi}_{\alpha,\epsilon} = \bar{\Psi}_\alpha \star \delta_\epsilon\\
\psi_{\alpha,\epsilon} = \psi_\alpha \star \delta_\epsilon\\
\bar{\psi}_{\alpha,\epsilon} = \bar{\psi}_\alpha \star \delta_\epsilon\\
H_{i,\epsilon} = H_i \star \delta_\epsilon\\
\bar{H}_{i,\epsilon} = H_i \star D_\epsilon\\
\eta_{i,\epsilon} = \eta_i \star \delta_\epsilon\\
\bar{\eta}_{i,\epsilon} = \bar{\eta}_i \star D_\epsilon.
\end{align*}

Define the cut-off Fermi action by

\begin{multline}\label{fermiaction}
{S}^{\lambda, \mu}_{F,\kappa,\epsilon} (H_i,\Psi_\alpha, \eta_i, \psi_\alpha,\bar{H}_i,\bar{\Psi}_\alpha, \bar{\eta}_i,\bar{\psi}_\alpha;\Phi)
=\\ S_{F,0} (H_i,\Psi_\alpha, \eta_i,{\psi}_\alpha,\bar{H}_i,\bar{\Psi}_\alpha, \bar{\eta}_i,\bar{\psi}_\alpha)+
S^{\lambda,\mu}_{F,I,\kappa} (H_{i,\epsilon},\Psi_{\alpha,\epsilon}, \eta_{i,\epsilon}, \psi_{\alpha,\epsilon},\bar{H}_{i,\epsilon},\bar{\Psi}_{\alpha,\epsilon}, \bar{\eta}_{i,\epsilon}, \bar{\psi}_{\alpha,\epsilon}
;\Phi).\end{multline}

The cut-off Fermi action ${S}^{\lambda, \mu}_{F,\kappa,\epsilon}$ gives rise to a formal perturbation series.  We will show in Section \ref{sec:fermionization} that in the limit $\epsilon \to 0,$ this series is equal, term-by-term, to the perturbation series of the Bosonic partition function $Z_\kappa(\lambda,\mu;\Phi).$  Thus

\begin{thm}\label{fermionization}  The action ${S}^{\lambda, \mu}_{F,\kappa,\epsilon}$ gives rise to a perturbation series (in $\lambda$), each term of which is equal in the limit $\epsilon \to 0$ to the corresponding term of the formal power series (\ref{pfdefeqn}).
\end{thm}

The proof of this theorem is a straightforward exercise in perturbation
theory, but like many perturbative calculations
it is very tedious to write down.  We postpone it to Section \ref{sec:fermionization}.

\begin{rem}  It is possible to write down a Fermi theory which gives rise to a perturbation series which is equal term-by-term to the formal power series (\ref{pfdefeqn}).  However, this theory does not arise from a Lagrangian; the free correlation functions for the Fermi fields $H_i, \eta_i$ in this theory are of the form
$$<\bar{\eta}_i(x) \eta_j(y)> = < \bar{H}_i(x) H_j(y) > = \delta_{ij} {\rm ~if~} x = y;$$
$$<\bar{\eta}_i(x) \eta_j(y)> = < \bar{H}_i(x) H_j(y) > = 0  {\rm ~if~} x \neq y,$$

\noindent as might be expected from the limiting behavior of the correlations of the cut-off fields $\bar{H}_{i,\epsilon}, H_{i,\epsilon},\bar{\eta}_{i,\epsilon},\eta_{i,\epsilon}.$

Morally, we have replaced the gauge field $A_\alpha^i(x)$ by the Fermi bilinear $\bar{H}_i(x) \bar{\Psi}_\alpha(x)$ and the conjugate field $F_\alpha^i(x)$ by $\bar{\eta}_i(x) \bar{\psi}_\alpha(x).$  In physical language the gauge bosons are formed from the Fermions by a process similar to the formation of Cooper pairs in superconductivity.  In that case the Fermions are physical electrons; I do not know whether in the case of gauge theory the Fermions have meaning as physical modes or whether they should be viewed as mathematical device.

It is interesting to note that the Fermionic theory we have constructed should morally give rise, in the limit $\kappa \to \infty,$ to a local, gauge invariant theory.  However, both locality and gauge invariance are not manifest in the Fermionic theory itself.  In essence we have exchanged manifest locality and gauge invariance for the ability to work with a Fermionic theory with good behavior of perturbation theory which is not manifest in the Bosonic formalism.\end{rem}

\subsection{Convergent perturbation theory.}  We have seen that Yang-Mills theory with spacetime and momentum cutoffs is equivalent to a nonlocal Fermionic theory.  However, it is well-known\footnote{The earliest reference I know of for this fact is \cite{caia}.} (see for example \cite{salmhofer}) that Fermionic quantum field theories with spacetime and momentum cutoffs give rise to perturbation series which are convergent.  This is in stark contrast to the case of Bosonic quantum field theories, where such convergence is not expected.\footnote{Indeed even the one-dimensional integral $z(\lambda) = \int_\R dx e^{-x^2 -\lambda x^4},$ which is well-defined for $\lambda > 0,$ does not have a convergent power series expansion at $\lambda=0.$}  The surprising result of this paper is that four dimensional Yang-Mills theory, when appropriately cut off, gives rise to a convergent power series expansion.  Thus the gauge fields, which are one-forms, have something of the flavor of Fermi fields.  

\begin{thm}\label{convergence} The cut-off Fermi action ${S}^{\lambda, \mu}_{F,\kappa,\epsilon}$ gives rise to a convergent perturbation series in $\lambda,$ with uniform bounds for convergence for $\mu \in \R$ and $\Phi \in \fg \otimes \R^3.$
\end{thm}

We give the proof of this theorem, which again is a straightforward exercise in perturbation theory, in Section \ref{sec:convergence}.

Thus the cut-off partition function 

$$Z_{F,\kappa,\epsilon}(\lambda, \mu;\Phi) := \int \cD \Psi \cD H \cD \psi \cD \eta
\cD\bar{\Psi} \cD\bar{H} \cD \bar{\psi} \cD \bar{\eta} e^{{S}^{\lambda, \mu}_{F,\kappa,\epsilon}(H_i,\Psi_\alpha, \eta_i, \psi_\alpha,\bar{H}_i,\bar{\Psi}_\alpha, \bar{\eta}_i,\bar{\psi}_\alpha ;\Phi)},$$

\noindent defined by its convergent perturbation series, is an analytic function of $\lambda$ for all $\mu \in \R$ and all $\Phi \in \fg \otimes \R^3.$

The Fermi action ${S}^{\lambda, \mu}_{F,\kappa,\epsilon}$ thus gives a description of quantum gauge theories by a convergent perturbation expansion.  This fact might explain the effectiveness of perturbation theory in predicting physical phenomena arising in gauge theories.  From the mathematical point of view, asymptotically free theories with convergent perturbation expansions can be constructed by classical methods of constructive quantum field theory; this was done for the Gross-Neveu model in two spacetime dimensions by Gawedzki and Kupiainen \cite{gk} and Feldman, Magnen, Rivasseau, and Seneor \cite{fmrs}.  Assuming that with the proper scaling of $\kappa$ and $\epsilon$ the novel cutoff in the definition of ${S}^{\lambda, \mu}_{F,\kappa,\epsilon}$ does not destroy the essential property of asymptotic freedom, we expect that our methods will yield a relatively simple method to construct the quantum Yang-Mills theory in four dimensions in finite volume, avoiding the difficulties associated with lattice methods.

We are also led to make the following
\begin{Conjecture} The formal power series (\ref{pfdefeqn}) is convergent for all $\lambda \in \C,$ with uniform bounds for convergence for
$\mu \in \R$ and $\Phi \in \fg \otimes \R^3.$
\end{Conjecture}

This conjecture is hardly crucial,
since the perturbation series for the Fermionic action $S^{\lambda,\mu}_{F,\kappa,\epsilon}$
with (say) $\epsilon \sim \kappa^{-a}$  where $a > 0$ is no less reasonable as a cut-off version of Yang-Mills theory than the perturbation series (\ref{pfdefeqn}).

\subsection{Decay of correlations.}
To take the infinite volume limit, we must study the behavior of correlations.  To zeroth order in perturbation theory, this is given by the behavior of the propagator $C.$  Let us consider the expectation

$$E(0,x) := \sum_{\alpha}< A^3_\alpha(0,0,0,0) A^3_\alpha (0,0,x,0)>$$

\noindent for transverse gauge modes as $x \to \infty.$  In the limit $\kappa \to \infty,$ we have, by equation
(\ref{apropagator}),
\begin{equation}\label{corrs}E(0,x) = \sum_{\alpha}\int_{\fg\otimes \R^3} d\Phi e^{-\frac12 |\Phi|^2} 
((\Delta_{\mu\Phi})^{-1}(0,0,0,0;0,0,x,0))_{\alpha,\alpha;3,3},\end{equation}
\noindent where the operator $\Delta_{\mu\Phi}$ is the linear operator on 
$\fg\otimes \R^3$-valued functions on the torus given by

$$\Delta_{\mu\Phi} = - \frac{d^2}{dx_0^2} + \nabla\times \nabla \times-2 i \mu [\Phi,\cdot].$$

In formula (\ref{corrs}) the infinite volume limit can be taken directly.  Given $\delta \in \fg,$ let $\fg_\delta:= {\rm ker~ ad}_\delta.$  For generic $\delta,$ $\fg_\delta$ is a maximal torus of $\fg,$ and in general $\fg_\delta$ always contains a maximal torus.  Consider the nonzero $\delta \in \fg$ such that $\fg_\delta$ is maximal.  All such maximal $\fg_\delta$'s are conjugate to one another, and hence to some fixed $\fg_{\delta_0} =: \tilde{\fg}.$  Let
$$d_G := {\rm dim~} \fg - {\rm dim~} \tilde{\fg}.$$

\begin{Conjecture}\label{decay}
(a)  In the limit $\Lambda_i\to \infty,$ $i=1,2,3,$ and $\kappa \to \infty,$ the correlation function $E(0,x)$ has the asymptotic behavior
\begin{equation}\label{decaybd}
E(0,x) \sim x^{-2 -2 d}
\end{equation}
\noindent as $x \to \infty$ for some positive integer $d.$
\newline
(b)  The integer $d$ appearing in (\ref{decaybd}) is given by $d=d_G.$
\end{Conjecture}

To make this conjecture plausible, we compute

\begin{multline}\label{approx}
\int_{\fg \otimes \R^3} d\Phi e^{-\frac12 |\Phi|^2} \left(\frac{1}{-\Delta -2  i \mu [\Phi,\cdot]}\right)^{\alpha,\alpha}_{3,3}(0,0,0,0;0,0,x,0)
=\\
\frac{1}{2} \int_{\fg \otimes \R^3} d\Phi e^{-\frac12 |\Phi|^2} \left(\frac{\Delta}{\Delta^2 + 4 \mu^2 [\Phi,\cdot]^2}\right)^{\alpha,\alpha}_{3,3}(0,0,0,0;0,0,x,0),
\end{multline}

\noindent where $\Delta$ is the ordinary Laplacian on $\R^4.$

While the expression (\ref{approx}) does not quite give the correct formula  (\ref{apropagator}) for the correlation function $E(0,x),$ it might be expected to give a rough idea of the behavior of $E(0,x).$  But from estimates on asymptotics of the propagator in four dimensions (see for example \cite{gj}), we expect that

$$
\left(\frac{\Delta}{\Delta^2 + 4 \mu^2 [\Phi,\cdot]^2}\right)^{\alpha,\alpha}_{3,3}(0,0,0,0;0,0,x,0)
\sim \frac{1}{x^2} e^{- 2 |\mu \eta|^{1/2} |x|}$$

\noindent where $\eta$ is the smallest eigenvalue of the mass matrix
$[\Phi, \cdot].$  A direct calculation for the case of $G=SU(2)$ shows that this mass matrix is nonsingular for generic $\Phi,$ but has a zero eigenvalue where two components of the $su(2)-$valued vector $\Phi$ commute.  In general it is easy to see that the mass matrix $[\Phi, \cdot]$ has a kernel when two components of the $\fg-$valued vector $\Phi,$ considered as operators on $\fg$ by the adjoint action, have kernels with nonzero intersection.  Given one such component $\gamma,$ this occurs precisely when another component lies in $\fg_\delta,$
where $\delta \in \fg_\gamma.$  Hence we expect exponential decay of correlations away from a subvariety of the space of $\Phi$'s of codimension $d_G.$  The integral over $\fg \otimes \R^3$ is dominated by the integral over a small neighborhood of this subvariety, and if we assume the lowest eigenvalue of the mass matrix is given by the distance from this subvariety, we obtain the bound of Conjecture \ref{decay}.  If we have not correctly described the subvariety of singular $[\Phi,\cdot]$'s, or if the behavior of the lowest eigenvalue of the mass matrix is not given by  the distance from this subvariety, but by some power of this distance, a polynomial bound of the form (\ref{decaybd}) should still hold with a different formula for $d.$

\begin{rem} We have exhibited the Fermionized Yang-Mills theory as a statistical ensemble of theories exhibiting a magnetic Higgs mechanism, but with the average dominated by theories with small mass.
Thus we do not expect four-dimensional gauge theory to possess a mass gap;\footnote{Of course it is possible that miraculous cancellations in higher orders of perturbation theory will result in more rapid decay, perhaps even exponential decay, of correlations.} however, the confinement mechanism of 't Hooft should still be effective, due to the rapid polynomial decay of correlations in the Euclidean theory.  Note that when $G=SU(n),$
$d_G = {\rm dim~}SU(n) - {\rm dim~}S(U(n-1) \times U(1)) = 2n -2$, so that
as $n \to \infty,$ the rate of decay increases, as expected from the study of gauge theories in the large $n$ limit. 
In mathematical terms, the decay of Conjecture \ref{decay} should still be
adequate to establish the existence of the infinite volume limit of this theory.  We also expect that if our cut-off theory is a reasonable description of the behavior of quantum Yang-Mills theory, the fact that this {\em Euclidean} theory exhibits polynomial, rather than exponential, decay of correlations should give rise to measurable effects in lattice gauge theory and perhaps in experimental settings.
\end{rem}
\begin{rem}
It would be useful to develop a better understanding of the structure of the mass term, given by the bracket $[\Phi,\cdot].$  This bracket arises in some sense from the hyperkahler structure on $T^4,$ and comes from the Poisson brackets of components of the three moment maps for the gauge group action, given by each of the three symplectic forms on the space of connections.  In some sense the space $\fg \otimes \R^3$ should be considered as a tensor product $\fg \otimes su(2),$ which is equipped with an {\em even} bracket, and should perhaps be considered as a Jordan algebra. Perhaps this kind of algebraic approach might be useful in proving that the mass term is generically nonsingular (which I was only able to do by direct calculation in the case of $G=SU(2)$) and in finding the subvariety corresponding to singular mass terms.\end{rem}
\section{Proof of Propositions \ref{propagator1} and \ref{propagator}}\label{section:propagator}

Since Proposition \ref{propagator1} follows from Proposition \ref{propagator} where $\Phi=0,$ we concentrate on Proposition \ref{propagator}.

To do this, we first note that since $A$ and $F$ have finite Fourier series,
the proof of Proposition \ref{propagator} is a problem in finite-dimensional
linear algebra.  Suppose the quadratic form given by $S_{0,\kappa}^\mu(\cdot,\cdot;\Phi)$ is degenerate.  Then there exists an element $(A,F) \in \cA^\kappa \times \cF_0^\kappa$ satisfying

\begin{align}
\langle F^\prime, (F - i \frac{dA}{dx_0} - i \nabla \times A)\rangle & = 0 \label{deg1}\\
\langle A^\prime, i \frac{dF}{dx_0} -i \nabla \times F -2 i \mu [\Phi,A]\rangle&  = 0\label{deg2}
\end{align}

\noindent for all $(A^\prime,F^\prime) \in \cA^\kappa \times \cF_0^\kappa.$

Thus, by (\ref{deg1})
\begin{equation}\label{Feqn}
F =  i \frac{dA}{dx_0} + i \nabla \times A + K
\end{equation}
\noindent where $K \in \fg \otimes \R^3$ is a constant.  Then by (\ref{deg2})
and (\ref{Feqn}),

$$||\frac{dA}{dx_0}||^2 + ||\nabla \times A||^2 -2 i \mu \langle A, [\Phi,A]\rangle = 0.$$

Since $\langle A, [\Phi,A]\rangle$ is real, it follows that $A= \nabla \phi + C,$ where $\phi$ is a $\fg$-valued function on the torus with $$\frac{d\phi}{dx_0} = 0,$$ and $C \in \fg \otimes \R^3$ is a constant.
If we consider $A$ as a periodic vector field on the cube   $D= [-\Lambda_0,\Lambda_0]\times [-\Lambda_1,\Lambda_1]\times [-\Lambda_2,\Lambda_2]\times [-\Lambda_3,\Lambda_3]\subset\R^4,$ and likewise we consider $\phi$ as a periodic function on $D,$
then $A = \nabla \tilde{\phi},$ where $\tilde{\phi} = \phi + \sum C_i x_i.$  The gauge conditions (\ref{gaugefix}) imply that $\tilde{\phi}$ is a constant, so that $A =0.$ By (\ref{Feqn}) $F$ must be a constant; since $F \in \cF_0^\kappa,$ $F=0$ as needed.

We can also obtain explicit expressions for the propagators 
${C}_{\mu,\kappa}^{A,A}(\Phi),$ ${C}_{\mu,\kappa}^{A,F}(\Phi),$ and ${C}_{\mu,\kappa}^{F,F}(\Phi).$  In particular,
\begin{equation}\label{apropagator}
{C}_{\mu,\kappa}^{A,A}(\Phi)=   (\Delta_{\mu\Phi})^{-1}\end{equation}
\noindent where $\Delta_{\mu\Phi}$ is the differential operator
$$\Delta_{\mu\Phi} = -\frac{d^2}{dx_0^2} + \nabla \times \nabla \times-2 i \mu[\Phi,\cdot].$$
\begin{rem}\label{ktoinf}In this paper we have not considered the limit $\kappa \to \infty,$ so the finite-dimensional linear algebraic formulas appearing in this section are adequate for our purposes. More care is required as $\kappa \to \infty.$  In this limit, the propagator ${C}^{F,F}_{\mu,\kappa}$ projected in the direction of the image of the de Rham operator $\nabla$ becomes a distribution supported on the diagonal  (as would the projection of ${C}^{F,F}_{\mu,\kappa}$ to the constants, had we not removed the constant mode from $F.$)  This may cause problems in Fermionization, since it may not be possible to remove the diagonals from the integrals appearing in the perturbation series (\ref{pfdefeqn}).
The projection of ${C}^{F,F}_{\mu,\kappa}$ onto the complement of this image, as well as the remaining components ${C}^{A,F}_{\mu,\kappa}$ and ${C}^{A,A}_{\mu,\kappa}$ of the propagator, become functions with algebraic singularities on the diagonal as $\kappa \to \infty.$
\end{rem}

\section{Proof of Theorem \ref{fermionization}}\label{sec:fermionization}

The formal perturbation series for the cut-off Yang-Mills theory in Hamiltonian form, as given by the action $S_{H,\kappa}^{\lambda, \mu}(A,F;\Phi),$ is given as in equation (\ref{pfdefeqn}) by 

\begin{multline}\label{bosept}
Z_\kappa(\lambda, \mu;\Phi) := \sum_{n=0}^\infty \Theta_n \lambda^{2n}=
\exp\Bigl  (\sum_{i,j,\alpha,\beta} \int_{T^4\times T^4} dx dy 
  \Bigl [
({C}_{\mu,\kappa}^{A,A}(\Phi)(x,y))_{i,j;\alpha,\beta} 
\frac{\delta}{\delta J^i_\alpha(x)}\frac{\delta}{\delta J^j_\beta(y)}\\+ 2({C}_{\mu,\kappa}^{A,F}(\Phi)(x,y))_{i,j;\alpha,\beta} 
\frac{\delta}{\delta J^i_\alpha(x)}\frac{\delta}{\delta K^j_\beta(y)}\\+
({C}_{\mu,\kappa}^{F,F}(\Phi)(x,y))_{i,j;\alpha,\beta} 
\frac{\delta}{\delta K^i_\alpha(x)}\frac{\delta}{\delta K^j_\beta(y)}
  \Bigr ] 
\Bigr )\\
\exp \left( i \lambda \sum_{i,j,k; \alpha,\beta,\gamma}
\epsilon_{ijk}
f_{\alpha\beta\gamma}\int_{T^4} dx J^i_\alpha(x) J^j_\beta(x) K^k_\gamma(x)
\right)_{J=K=0}
\end{multline}

The perturbation series of the Fermi action $S^{\lambda, \mu}_{F,\kappa,\epsilon}$ is given by\footnote{Recall we are expanding in $\lambda$ with $\mu$ and $\Phi$ held constant.}

\begin{equation}\label{fermipt}
Z_{F,\kappa,\epsilon}(\lambda, \mu;\Phi) := \sum_{n=0}^\infty \Xi_n(\epsilon) \lambda^{2n}\end{equation}

\noindent where
\begin{multline*}
\Xi_n(\epsilon) = \frac{1}{(2n)!(3n)!}\int \cD \Psi \cD H \cD \psi \cD \eta
\cD\bar{\Psi} \cD\bar{H} \cD \bar{\psi} \cD \bar{\eta}\\
e^{S_{F,0}}
\Bigl (
(-i/2) \int_{T^4} dx \sum_{\alpha,\beta,\gamma;i,j,k} \epsilon_{ijk}
f_{\alpha\beta\gamma}  
\bar{H}_{i,\epsilon}(x) 
\bar{\Psi}_{\alpha,\epsilon}(x) 
\bar{H}_{j,\epsilon}(x) 
\bar{\Psi}_{\beta,\epsilon}(x)
\bar{\eta}_{k,\epsilon}(x)
\bar{\psi}_{\gamma,\epsilon}(x)
\Bigr )^{2n}\\
\Bigl ( \int_{T^4 \times T^4} dx dy \Bigl [\sum_{\alpha,\beta,i,j}
H_{i,\epsilon}(x) \Psi_{\alpha,\epsilon}(x) ({C}_{\mu,\kappa}^{A,A}(\Phi)(x,y))_{\alpha,\beta; i,j} H_{j,\epsilon}(y) \Psi_{\beta,\epsilon}(y) +\\
2 H_{i,\epsilon}(x) \Psi_{\alpha,\epsilon}(x) ({C}_{\mu,\kappa}^{A,F}(\Phi)(x,y))_{\alpha,\beta; i,j}
 \eta_{j,\epsilon}(y) \psi_{\beta,\epsilon}(y) +\\
\eta_{i,\epsilon}(x) \psi_{\alpha,\epsilon}(x) ({C}_{\mu,\kappa}^{F,F}(\Phi)(x,y))_{\alpha,\beta; i,j\
} \eta_{j,\epsilon}(y) \psi_{\beta,\epsilon}(y)\Bigr ]
\Bigr )^{3n}.
\end{multline*}

Standard Feynman diagram techniques allow us to write $\Xi_n(\epsilon)$ as
a sum of terms, each of which corresponds to a pair of trivalent diagrams.  This pair consists of a diagram given by the pairings of the $\Psi, \psi, \bar{\Psi},$ and $\bar{\psi}$ fields, and another diagram given by the pairings of the $H, \eta, \bar{H},$ and $\bar{\eta}$ fields.  We use this fact to write
$$ \Xi_n(\epsilon) = \Xi_n^1(\epsilon) + \Xi_n^2 (\epsilon)$$

\noindent where $\Xi_n^1(\epsilon)$ is the sum of those terms in the
diagrammatic expansion of $\Xi_n(\epsilon)$ corresponding to pairs of
{\em identical} Feynman diagrams---that is, Feynman diagrams where the
combinatorics of the pairings of the $\Psi, \psi, \bar{\Psi},$
and $\bar{\psi}$ fields are the same as those of the $H, \eta, \bar{H},$ and $\bar{\eta}$ fields; in the limit $\epsilon \to 0$ we will see that the terms
appearing in $\Xi_n^1(\epsilon)$ approach the corresponding terms in
(\ref{bosept}).
The sum of the remaining terms in the diagrammatic expansion of $\Xi_n(\epsilon),$ which we denote by $\Xi_n^2(\epsilon),$
consists of terms corresponding to pairs of Feynman diagrams where at least
one $H$ or $\eta$ propagator is not matched by a corresponding $\Psi$ or
$\psi$ propagator.  We will see that in the limit $\epsilon \to 0,$ such an
``unmatched''  propagator will give rise to a factor of order
$O(\epsilon^4),$ so that $\lim_{\epsilon\to 0} \Xi_n^2(\epsilon) = 0.$
More precisely, we write
\begin{multline}\label{xi1}
\Xi_n^1(\epsilon) := \frac{1}{(2n)!(3n)!}\Bigl (-\frac{i}{2}\Bigr)^{2n}\\
\sum_{\sigma\in S_{6n}}\int_{{(T^4)}^{6n}} dz_1\cdots dz_{6n} 
\int_{{(T^4)}^{6n}} dx_1\cdots dx_{6n}
\sum_{k=1}^{6n} \sum_{A_k \in \{A,F\}}\sum_{\alpha_k,\beta_k=1,\cdots, {\rm dim~}\fg} \sum_{i_k,j_k=1,2,3}\\
\prod_{m=1}^{2n} \delta(z_{3m}-z_{3m-1})\delta(z_{3m}- z_{3m-2})
\prod_{l=1}^{3n} (C^{A_{2l},A_{2l-1}}_{\mu,\kappa}(x_{2l},x_{2l-1}))_{i_{2l},i_{2l-1}; \alpha_{2l}, \alpha_{2l-1}}
\prod_{m=1}^{2n} 
\epsilon_{j_{3m}j_{3m-1}j_{3m-2}} f_{\beta_{3m}\beta_{3m-1}\beta_{3m-2}}\\
\prod_{m=1}^{2n} \frac12\Bigl(
\int \cD \Psi \cD H \cD \psi \cD \eta
\cD\bar{\Psi} \cD\bar{H} \cD \bar{\psi} \cD \bar{\eta}e^{S_{F,0}}
H^{A_{\sigma(3m)}}_{i_{\sigma(3m)},\epsilon}(x_{\sigma(3m)})
\Psi^{A_{\sigma(3m)}}_{\alpha_{\sigma(3m)},\epsilon}(x_{\sigma(3m)})\\
H^{A_{\sigma(3m-1)}}_{i_{\sigma(3m-1)},\epsilon}(x_{\sigma(3m-1)})
\Psi^{A_{\sigma(3m-1)}}_{\alpha_{\sigma(3m-1)},\epsilon}(x_{\sigma(3m-1)})
H^{A_{\sigma(3m-2)}}_{i_{\sigma(3m-2)},\epsilon}(x_{\sigma(3m-2)})
\Psi^{A_{\sigma(3m-2)}}_{\alpha_{\sigma(3m-2)},\epsilon}(x_{\sigma(3m-2)})\\
\bar{H}_{j_{3m},\epsilon}(z_{3m})
\bar{\Psi}_{\beta_{3m},\epsilon}(z_{3m})
\bar{H}_{j_{3m-1},\epsilon}(z_{3m-1})
\bar{\Psi}_{\beta_{3m-1},\epsilon}(z_{3m-1})
\bar{\eta}_{j_{3m-2},\epsilon}(z_{3m-2})
\bar{\psi}_{\beta_{3m-2},\epsilon}(z_{3m-2})\Bigr).
\end{multline}

Here the variables $A_l$ take values in the set of letters $\{A,F\},$ with the convention that $H_{i,\epsilon}^A = H_{i,\epsilon},$ $H_{i,\epsilon}^F = \eta_{i,\epsilon},$ $\Psi_{\alpha,\epsilon}^A = \Psi_{\alpha,\epsilon},$ $\Psi_{\alpha,\epsilon}^F = \psi_{\alpha,\epsilon},$ and similarly for complex conjugates.
We write $C^{F,A}_{\mu,\kappa}(\Phi)(x,y) := C^{A,F}_{\mu,\kappa}(\Phi)(y,x),$ and  $\delta_{A,A}=\delta_{A,F}=1, \delta_{A,F}= \delta_{F,A}= 0.$ 

To evaluate $\Xi_n^2(\epsilon),$ we use the following notation.  Given $\sigma,\tau \in S_{6n},$ we say $\tau \sim \sigma,$ if for every $m =1,\dots, 6n,$ there exists $k \in 1,\dots, n$ such that $\sigma(m), \tau(m) \in \{3k,3k+1,3k+2\}.$  If $\sigma \nsim \tau,$ let $m=m(\sigma,\tau)$ be the smallest integer so that 
$\sigma(m), \tau(m) \notin \{3k,3k+1,3k+2\}$ for any $k.$  Then

\begin{multline}\label{xi2}
\Xi_n^2(\epsilon) :=\frac{1}{(2n)!(3n)!}\Bigl(-\frac{i}{2}\Bigr)^{2n}\\
\sum_{\sigma\in S_{6n}}
\sum_{\tau \in S_{6n}, \tau \nsim \sigma}
{\rm sgn}(\sigma,\tau)
\int_{(T^4)^{6n}} dz_1\cdots dz_{6n} 
\int_{(T^4)^{6n}} dx_1\cdots dx_{6n}\\
\sum_{k=1}^{6n} \sum_{A_k \in \{A,F\}}\sum_{\alpha_k,\beta_k=1,\cdots, {\rm dim~}\fg} \sum_{i_k,j_k=1,2,3}
\Bigl(\prod_{m=1}^{2n} \delta(z_{3m}-z_{3m-1})\delta(z_{3m}- z_{3m-2})\Bigr)\\
\Bigl(\prod_{l=1}^{3n} (C^{A_{2l},A_{2l-1}}_{\mu,\kappa}(x_{2l},x_{2l-1}))_{i_{2l},i_{2l-1}; \alpha_{2l}, \alpha_{2l-1}}\Bigr)
\Bigl(\prod_{m=1}^{2n} \epsilon_{j_{3m}j_{3m-1}j_{3m-2}} f_{\beta_{3m}\beta_{3m-1}\beta_{3m-2}}\Bigr)\\
\Bigl(\prod_{\stackrel{l=1}{l\equiv 0,1(3)}}^{6n} \delta_{A_{\sigma(l)},A}\delta_{A_{\tau(l)},A}\Bigr)
\Bigl(\prod_{\stackrel{l=1}{l\equiv 2(3)}}^{6n} \delta_{A_{\sigma(l)},F}\delta_{A_{\tau(l)},F}\Bigr)
\Bigl(\prod_{l=1}^{6n} \delta_{\alpha_{\sigma(l)},\beta_l}\delta_{i_{\tau(l)},j_l}\Bigr)\\
\Bigl(\prod_{\stackrel{l=1}{l\neq m(\sigma,\tau), \sigma^{-1}\circ \tau(m(\sigma,\tau))}}^{6n}
\tilde{\delta}_\epsilon(x_{\sigma(l)} -z_l)
\tilde{D}_\epsilon(x_{\tau(l)} -z_l)\Bigr)\\
\tilde{\delta}_\epsilon(x_{\sigma(m(\sigma,\tau))}-z_{m(\sigma,\tau)})
\tilde{D}_\epsilon(x_{\tau(m(\sigma,\tau))}-z_{m(\sigma,\tau)})
\tilde{\delta}_\epsilon(x_{\tau(m(\sigma,\tau))}-z_{\sigma^{-1}\circ\tau(m(\sigma,\tau))})
\tilde{D}_\epsilon(x_{\tau\circ{\sigma^{-1}}\circ \tau(m(\sigma,\tau))}-
z_{\sigma^{-1}\circ \tau(m(\sigma,\tau))}).
\end{multline}

Here ${\rm sgn}(\sigma,\tau)$ is an element of $\{\pm 1\}$ which we will not need to compute explicitly, and we have used the notation

$$\tilde{\delta}_\epsilon = \delta_\epsilon \star \delta_\epsilon$$

\noindent and

$$\tilde{D}_\epsilon = \delta_\epsilon \star D_\epsilon.$$

We now proceed to evaluate $\Xi_n^1(\epsilon).$  We first note that
\begin{multline*}
\Bigl(-\frac{i}{4}\Bigr)
\sum_{j_{3m},j_{3m-1},j_{3m-2},\beta_{3m},\beta_{3m-1},\beta_{3m-2}}
\epsilon_{j_{3m}j_{3m-1}j_{3m-2}} f_{\beta_{3m}\beta_{3m-1}\beta_{3m-2}}\\
\delta(z_{3m}-z_{3m-1})\delta(z_{3m}- z_{3m-2})
\int \cD \Psi \cD H \cD \psi \cD \eta
\cD\bar{\Psi} \cD\bar{H} \cD \bar{\psi} \cD \bar{\eta}e^{S_{F,0}}
H^{A_{\sigma(3m)}}_{i_{\sigma(3m)},\epsilon}(x_{\sigma(3m)})
\Psi^{A_{\sigma(3m)}}_{\alpha_{\sigma(3m)},\epsilon}(x_{\sigma(3m)})
\\
H^{A_{\sigma(3m-1)}}_{i_{\sigma(3m-1)},\epsilon}(x_{\sigma(3m-1)})
\Psi^{A_{\sigma(3m-1)}}_{\alpha_{\sigma(3m-1)},\epsilon}(x_{\sigma(3m-1)})
H^{A_{\sigma(3m-2)}}_{i_{\sigma(3m-2)},\epsilon}(x_{\sigma(3m-2)})
\Psi^{A_{\sigma(3m-2)}}_{\alpha_{\sigma(3m-2)},\epsilon}(x_{\sigma(3m-2)})\\
\bar{H}_{j_{3m},\epsilon}(z_{3m})
\bar{\Psi}_{\beta_{3m},\epsilon}(z_{3m})
\bar{H}_{j_{3m-1},\epsilon}(z_{3m-1})
\bar{\Psi}_{\beta_{3m-1},\epsilon}(z_{3m-1})
\bar{\eta}_{j_{3m-2},\epsilon}(z_{3m-2})
\bar{\psi}_{\beta_{3m-2},\epsilon}(z_{3m-2})=\\
\frac{i}2
\epsilon_{i_{\sigma(3m)}i_{\sigma(3m-1)}i_{\sigma(3m-2)}}
f_{\alpha_{\sigma(3m)}\alpha_{\sigma(3m-1)}\alpha_{\sigma(3m-2)}}\\
\Bigl (\delta_{A_{\sigma(3m)},A}\delta_{A_{\sigma(3m-1)},A}\delta_{A_{\sigma(3m-2)},F}
+\delta_{A_{\sigma(3m)},A}\delta_{A_{\sigma(3m-1)},F}\delta_{A_{\sigma(3m-2)},A}
+\delta_{A_{\sigma(3m)},F}\delta_{A_{\sigma(3m-1)},A}\delta_{A_{\sigma(3m-2)},A}
\Bigr )\\
\tilde{\delta}_\epsilon(x_{\sigma(3m)}-z_{3m})
\tilde{D}_\epsilon(x_{\sigma(3m)}-z_{3m})
\tilde{\delta}_\epsilon(x_{\sigma(3m-1)}-z_{3m})
\tilde{D}_\epsilon(x_{\sigma(3m-1)}-z_{3m})
\tilde{\delta}_\epsilon(x_{\sigma(3m-2)}-z_{3m})
\tilde{D}_\epsilon(x_{\sigma(3m-2)}-z_{3m}).
\end{multline*}

\begin{Lemma}\label{deltas}
The functions $\tilde{D}_\epsilon$ and $\tilde{\delta}_\epsilon$ are positive.  Furthermore
\begin{itemize}
\item $\lim_{\epsilon\to 0}\tilde{\delta}_\epsilon = \delta$ (in  ${\mathcal D'}(T^4)$).
\item $\tilde{\delta}_\epsilon \tilde{D}_\epsilon = \tilde{\delta}_\epsilon.$
\item $||\tilde{D}_\epsilon\star \tilde{\delta}_\epsilon||_1 = O(\epsilon^4).$
\item $||\tilde{\delta}_\epsilon||_1 = 1.$
\item $||\tilde{D}_\epsilon||_\infty = 1.$
\end{itemize}
\end{Lemma}

Applying Lemma \ref{deltas}, we see that

\begin{multline*}
 \epsilon_{i_{\sigma(3m)}i_{\sigma(3m-1)}i_{\sigma(3m-2)}}
f_{\alpha_{\sigma(3m)}\alpha_{\sigma(3m-1)}\alpha_{\sigma(3m-2)}}\\
\Bigl (\delta_{A_{\sigma(3m)},A}\delta_{A_{\sigma(3m-1)},A}\delta_{A_{\sigma(3m-2)},F}
+\delta_{A_{\sigma(3m)},A}\delta_{A_{\sigma(3m-1)},F}\delta_{A_{\sigma(3m-2)},A}
+\delta_{A_{\sigma(3m)},F}\delta_{A_{\sigma(3m-1)},A}\delta_{A_{\sigma(3m-2)},A}
\Bigr )\\
\lim_{\epsilon\to 0}
\tilde{\delta}_\epsilon(x_{\sigma(3m)}-z_{3m})
\tilde{D}_\epsilon(x_{\sigma(3m)}-z_{3m})
\tilde{\delta}_\epsilon(x_{\sigma(3m-1)}-z_{3m})
\tilde{D}_\epsilon(x_{\sigma(3m-1)}-z_{3m})
\tilde{\delta}_\epsilon(x_{\sigma(3m-2)}-z_{3m})
\tilde{D}_\epsilon(x_{\sigma(3m-2)}-z_{3m})
= \\
\sum_{j_{3m},j_{3m-1},j_{3m-2}}\sum_{\beta_{3m},\beta_{3m-1},\beta_{3m-2}}
\epsilon_{j_{3m}j_{3m-1}j_{3m-2}} f_{\beta_{3m}\beta_{3m-1}\beta_{3m-2}}\\
\frac{\delta}{\delta J^{A_{\sigma(3m)},i_{\sigma(3m)}}_{\alpha_{\sigma(3m)}} (x_{\sigma(3m)})  }
\frac{\delta}{\delta J^{A_{\sigma(3m-1)},i_{\sigma(3m-1)}}_{\alpha_{\sigma(3m-1)}}(x_{\sigma(3m-1)})}
\frac{\delta}{\delta J^{A_{\sigma(3m-2)},i_{\sigma(3m-2)}}_{\alpha_{\sigma(3m-2)}}(x_{\sigma(3m-2)})}\\
 J^{j_{3m}}_{\beta_{3m}}(z_{3m}) J^{j_{3m-1}}_{\beta_{3m-1}}(z_{3m}) K^{j_{3m-2}}_{\alpha_{3m-2}}(z_{3m}) \Big\vert_{J=K=0},
\end{multline*}

\noindent where we have used the notation 

$$J^{A,i}_\alpha := J^i_\alpha$$
$$J^{F,i}_\alpha := K^i_\alpha.$$

Thus 

\begin{multline*}
\lim_{\epsilon \to 0} \Xi_n^1(\epsilon) =\\
\frac{1}{(2n)!(3n)!}i^{2n}
\sum_{\sigma\in S_{6n}}\int_{{(T^4)}^{6n}} dz_1\cdots dz_{6n} 
\int_{{(T^4)}^{6n}} dx_1\cdots dx_{6n}
\sum_{k=1}^{6n} \sum_{A_k \in \{A,F\}}\sum_{\alpha_k,\beta_k=1,\cdots, {\rm dim~}\fg} \sum_{i_k,j_k=1,2,3}\\
\prod_{m=1}^{2n} \delta(z_{3m}-z_{3m-1})\delta(z_{3m}- z_{3m-2})
\prod_{l=1}^{3n} (C^{A_{2l},A_{2l-1}}_{\mu,\kappa}(x_{2l},x_{2l-1}))_{i_{2l},i_{2l-1}; \alpha_{2l}, \alpha_{2l-1}}\\
\prod_{m=1}^{2n} 
\epsilon_{j_{3m}j_{3m-1}j_{3m-2}} f_{\beta_{3m}\beta_{3m-1}\beta_{3m-2}}\\
\prod_{m=1}^{2n} \frac{1}{2}\Bigl(
\frac{\delta}{\delta J^{A_{\sigma(3m)},i_{\sigma(3m)}}_{\alpha_{\sigma(3m)}}(x_{\sigma(3m)})}
\frac{\delta}{\delta J^{A_{\sigma(3m-1)},i_{\sigma(3m-1)}}_{\alpha_{\sigma(3m-1)}}(x_{\sigma(3m-1)})}
\frac{\delta}{\delta J^{A_{\sigma(3m-2)},i_{\sigma(3m-2)}}_{\alpha_{\sigma(3m-2)}}(x_{\sigma(3m-2)})}\\
J^{j_{3m}}_{\beta_{3m}}(z_{3m})
J^{j_{3m-1}}_{\beta_{3m-1}}(z_{3m-1})
K^{j_{3m-2}}_{\beta_{3m-2}}(z_{3m-2})\Bigr)_{J=K=0}.
\end{multline*}

\noindent Comparing to (\ref{bosept}), we see therefore that

$$\lim_{\epsilon\to 0 } \Xi_n^1(\epsilon) = \Theta_n$$

\noindent The proof of Theorem \ref{fermionization} will be completed by the following

\begin{Lemma}
The term $\Xi_n^2(\epsilon)$ satisfies

$$\lim_{\epsilon \to 0} \Xi_n^2(\epsilon) = 0.$$
\end{Lemma}

To prove this, we examine the explicit expression (\ref{xi2}).  We have, up to a constant $K$ independent of $\epsilon,$

$$|\Xi_n^2(\epsilon)| \leq K \sup_{i,j,\alpha,\beta,A,A'} 
||(C^{A,A'}_{\mu,\kappa}(\cdot,\cdot))_{i,j;\alpha,\beta}||_\infty^{3n}
||\tilde{\delta}_\epsilon||_1^{6n-1} ||\tilde{D}_\epsilon||_\infty^{6n-1} 
||\tilde{D}_\epsilon \star \tilde{\delta}_\epsilon||_1.$$

\noindent But by Lemma \ref{deltas},
$$||\tilde{D}_\epsilon \star \tilde{\delta}_\epsilon||_1 = O(\epsilon^4),$$
\noindent while
$$||\tilde{\delta}_\epsilon||_1 = 1,$$
\noindent and
$$||\tilde{D}_\epsilon||_\infty = 1.$$
\noindent

Hence 
$$|\Xi_n^2(\epsilon)| = O(\epsilon^4)$$

\noindent as needed.

\section{Proof of Theorem \ref{convergence}.}\label{sec:convergence}

We first write down an explicit expression for the terms in the perturbation series of the action ${S}^{\lambda, \mu}_{F,\kappa,\epsilon}.$  The term $\Xi_n(\epsilon)$ of order $\lambda^{2n}$ in this expansion is given by the Berezin integral

\begin{multline*}
\Xi_n(\epsilon) = \frac{1}{(2n)!(3n)!}\int \cD \Psi \cD H \cD \psi \cD \eta
\cD\bar{\Psi} \cD\bar{H} \cD \bar{\psi} \cD \bar{\eta}\\
e^{S_{F,0} (H_i,\Psi_\alpha, \eta_i, \psi_\alpha)}
\Bigl (
(-i/2) \lambda \int_{T^4} dx \sum_{\alpha,\beta,\gamma;i,j,k} \epsilon_{ijk}
f_{\alpha\beta\gamma}  
\bar{H}_{i,\epsilon}(x) 
\bar{\Psi}_{\alpha,\epsilon}(x) 
\bar{H}_{j,\epsilon}(x) 
\bar{\Psi}_{\beta,\epsilon}(x) 
\bar{\eta}_{k,\epsilon}(x)
\bar{\psi}_{\gamma,\epsilon}(x)
\Bigr )^{2n}\\
\Bigl ( \int_{T^4 \times T^4} dx dy \Bigl [\sum_{\alpha,\beta,i,j}
H_{i,\epsilon}(x) \Psi_{\alpha,\epsilon}(x) ({C}_{\mu,\kappa}^{A,A}(\Phi)(x,y))_{\alpha,\beta; i,j} H_{j,\epsilon}(y) \Psi_{\beta,\epsilon}(y) +\\
2 H_{i,\epsilon}(x) \Psi_{\alpha,\epsilon}(x) ({C}_{\mu,\kappa}^{A,F}(\Phi)(x,y))_{\alpha,\beta; i,j}
 \eta_{j,\epsilon}(y) \psi_{\beta,\epsilon}(y) +\\
\eta_{i,\epsilon}(x) \psi_{\alpha,\epsilon}(x) ({C}_{\mu,\kappa}^{F,F}(\Phi)(x,y))_{\alpha,\beta; i,j\
} \eta_{j,\epsilon}(y) \psi_{\beta,\epsilon}(y)\Bigr ]
\Bigr )^{3n}
\end{multline*}

Theorem \ref{convergence} follows from the following estimate.

\begin{prop}\label{estimate}
There exists a constant $C>0$ (independent of $\Phi$ and $\mu$ for $\Phi \in \fg \otimes \R^3$ and $\mu \in \R$) such that

$$|\Xi_n(\epsilon)| < \frac{C^n}{(2n)!(3n)!}.$$
\end{prop}

To prove Proposition \ref{estimate}, we note that $\Xi_n(\epsilon)$ is a sum of $O(C^n)$ terms, each of which is (up to a constant of order $O(C^n)$) of the form

\begin{multline}\label{term}
\frac{1}{(2n)!(3n)!}\int \cD \Psi \cD H \cD \psi \cD \eta \cD\bar{\Psi} \cD\bar{H} \cD \bar{\psi} \cD \bar{\eta}\exp (\int_{T^4} dx \sum_i |H_i(x)|^2 + \sum_i |\eta_i(x)|^2+\sum_\alpha |\Psi_\alpha(x)|^2 + \sum_\alpha |\psi_\alpha(x)|^2)\\
\int_{(T^4)^{6n}} dx_1 dy_1 \dots dx_{3n} dy_{3n} \prod_{l=1}^{3n}
({C}_{\mu,\kappa}^{a_l,b_l}(\Phi)(x_l,y_l))_{i_l,j_l;\alpha_l,\beta_l}
H_{i_l,\epsilon}^{a_l}(x_l) \Psi_{\alpha_l,\epsilon}^{a_l}(x_l) H_{j_l,\epsilon}^{b_l}(y_l) \Psi_{\beta_l,\epsilon}^{b_l}(y_l)
\\
\int dz_1\dots dz_{2n} \prod_{m=1}^{2n} f_{\theta_m\iota_m\sigma_m} \epsilon_{i_mj_mk_m} 
\bar{H}_{p_m,\epsilon}(z_m) 
\bar{\Psi}_{\theta_m,\epsilon}(z_m) 
\bar{H}_{q_m,\epsilon}(z_m) 
\bar{\Psi}_{\iota_m,\epsilon}(z_m) 
\bar{\eta}_{r_m,\epsilon}(z_m)
\bar{\psi}_{\sigma_m,\epsilon}(z_m).
\end{multline}

Here the variables $a_m, b_m$ take values in the set of letters $\{A,F\}$ and as before we use the convention $H_{i,\epsilon}^A = H_{i,\epsilon},$ $H_{i,\epsilon}^F = \eta_{i,\epsilon},$ $\Psi_{\alpha,\epsilon}^A = \Psi_{\alpha,\epsilon},$ $\Psi_{\alpha,\epsilon}^F = \psi_{\alpha,\epsilon},$ with similar notation for complex conjugates, and we write $C^{F,A}_{\mu,\kappa}(\Phi)(x,y) :=
C^{A,F}_{\mu,\kappa}(\Phi)(y,x).$
The Berezin integral appearing in (\ref{term}) is
\begin{multline*}
B:=\int \cD \Psi \cD H \cD \psi \cD \eta \cD\bar{\Psi} \cD\bar{H} \cD \bar{\psi} \cD \bar{\eta}\exp (\int_{T^4} dx \sum_i |H_i(x)|^2 + \sum_i |\eta_i(x)|^2+\sum_\alpha |\Psi_\alpha(x)|^2 + \sum_\alpha |\psi_\alpha(x)|^2)\\
\prod_{l=1}^{3n}
H_{i_l,\epsilon}^{a_l}(x_l) \Psi_{\alpha_l,\epsilon}^{a_l}(x_l) H_{j_l,\epsilon}^{b_l}(y_l) \Psi_{\beta_l,\epsilon}^{b_l}(y_l)\\
\prod_{m=1}^{2n}
\bar{H}_{p_m,\epsilon}(z_m) 
\bar{\Psi}_{\theta_m,\epsilon}(z_m) 
\bar{H}_{q_m,\epsilon}(z_m) 
\bar{\Psi}_{\iota_m,\epsilon}(z_m) 
\bar{\eta}_{r_m,\epsilon}(z_m)
\bar{\psi}_{\sigma_m,\epsilon}(z_m).
\end{multline*}

This integral is the inner product of two elements of
$\bigwedge^{12n} (L_2(T^4)\otimes (\R^3\oplus \fg)\otimes \R^2),$ and is bounded by 

$$|B| < ||\delta_\epsilon||_{L_2(T^4)}^{18n} ||D_\epsilon||_{L_2(T^4)}^{6n}.$$

\noindent (see for example \cite{salmhofer}, p.199).

Since the kernels $({C}_{\mu,\kappa}^{a,b}(\Phi)(\cdot,\cdot))_{i,j;\alpha,\beta}$  have $L_\infty$ bounds uniform in $\mu$ and $\Phi$ by Proposition \ref{propagator}, Proposition \ref{estimate} follows.

\section{Remarks}
\subsection{Correlation functions, change of gauge and cutoff, QCD, supersymmetric theories.}  We have shown that the partition function of four-dimensional Yang-Mills theory in axial gauge with cutoffs can be written as the partition function of a Fermionic theory, which with a further cut-off has a convergent perturbation expansion.  By adding a source term $\int \sum_{i,\alpha} J^i_\alpha(x) A^i_\alpha(x)$ to the gauge action,
and a similar term $\int \sum_{i,\alpha} J^i_\alpha(x) \bar{H}_i(x) \bar{\Psi}_\alpha(x)$ to the Fermionized action, the same theorem can be proved for correlation functions.  Likewise the choice of axial gauge is inessential; the same result holds for covariant gauges, with the appropriate ghost terms included.  (In three dimensions we show how to deal with such terms in \cite{cspaper}.)
The ultraviolet and spacial cutoffs were chosen for technical convenience; since the theory turns out to be essentially perturbative, there should be no difficulty in replacing these cutoffs by other schemes, such as dimensional regularization, which may be more convenient.  Similarly, theories coupled to Fermionic matter fields can be Fermionized by the same method, and again perturbation theory will be convergent once an appropriate cutoff is introduced.  Similar methods should work for supersymmetric Yang-Mills theories containing only gauge and Fermion fields, whether on $T^4$ or on other manifolds.  On the other hand, in Yang-Mills-Higgs models (such as supersymmetric models containing a Higgs field), fermionization of the gauge field will go through, but we do not expect that these methods will yield a convergent perturbation expansion in the presence of a Bosonic Higgs field.  A construction of these models will therefore entail a combination of a perturbative treatment of the gauge field with nonperturbative methods appropriate to polynomial interactions of scalar fields.

\subsection{Instantons and winding modes.}\label{instanton}

We have only considered gauge theory on the trivial bundle, which excludes the possible presence of states with nonzero instanton number.  This is easy to remedy by replacing the trivial bundle with a nontrivial bundle $P\to T^4,$ choosing some connection $A_0$ on $P,$ and noting that any other connection on $P$ differs from $A_0$ by an element of $\Omega^1(T^4,ad(P)).$  We obtain an expression for the Yang-Mills action which is a slight modification of (\ref{yma}) and which can be treated by similar methods.  A similar shift in $A$ can be used to introduce winding modes.

\subsection{The limit ${\epsilon} \to 0, \kappa \to \infty$}  The convergence estimate of Proposition \ref{estimate} is not uniform in $\epsilon$ or $\kappa,$ and indeed as in \cite{fmrs,gk} we do not expect the renormalized perturbation series to be convergent.  However, in an asymptotically free theory, the coupling constant approaches zero as the cutoff is increased.  It is this fact that underlies the methods of \cite{fmrs,gk}, where theories with similar bounds for the perturbation series of the regularized theory, also not uniform in the cutoff, were constructed.\footnote{A rough calculation shows that if we take $\epsilon = \kappa^{-a},$ where $a > 2,$ the divergence structure of the theory should be the same as that of ordinary cut-off Yang-Mills; the spurious terms in the Feynman graph expansion encountered as $\Xi_n^2(\epsilon)$ in Section \ref{sec:fermionization} are all finite as $\kappa \to \infty.$}
It remains to be seen if the methods of \cite{fmrs,gk} apply to the case of Yang-Mills theory.

\subsection{Perturbation theory in $\lambda^{-1/3}$ }
The form of the Fermionic action $S^{\lambda, \mu}_{F,\kappa,\epsilon}$ suggests that there may exist a well-defined perturbation series in $1/\lambda,$ which may be worth studying.  Let us consider a toy model where the integrals appearing in the definitions of $S_{F,0}$ and $S_{F,I,\kappa}^{\lambda,\mu}$ are replaced by sums over a finite number of points.  In this case the Berezin integral 
$$Z:=\int \cD \Psi \cD H \cD \psi \cD \eta
\cD\bar{\Psi} \cD\bar{H} \cD \bar{\psi} \cD \bar{\eta} e^{S^{\lambda, \mu}_{F,I,\kappa}}$$
\noindent becomes a finite-dimensional determinant; we can imagine now computing the perturbation series 

$$\frac{1}{Z} \int \cD \Psi \cD H \cD \psi \cD \eta
\cD\bar{\Psi} \cD\bar{H} \cD \bar{\psi} \cD \bar{\eta} e^{S^{\lambda, \mu}_{F,I,\kappa}} e^{\beta S_{F,0}}$$

\noindent as a (finite) power series in $\beta.$  A rescaling shows that this is equivalent to a perturbation series in $\lambda^{-1/3},$ and it is immediately obvious that the correlation function $<A_i^\alpha(x) A_j^\gamma(y)>$ vanishes identically for all $x,y$ to lowest order in $\lambda^{-1/3}.$  This is suggestive of a strong coupling expansion where the correlations vanish---not merely decay---to lowest order.  I do not know how to extend these ideas beyond the toy model case or whether they are helpful at all.

\subsection{Direct Minkowski Construction.} Since the Fermionic theory we are considering is purely perturbative, it may be amenable to a direct construction in Minkowski space.  Of course the behavior of the propagators is much more subtle in the Minkowski setting, so such a construction would entail more delicate analysis than the methods used here.


\begin{thebibliography}{asdf}
%

\bibitem{caia} E. Caianello, {\em Nuovo Cimento} {\bf 3,} 223-225 (1956)
\bibitem{coleman} S. Coleman.  Aspects of Symmetry.  Cambridge University Press, 1985
\bibitem{c1} S. Coleman, Phys. Rev. D 11, 2088 - 2097 (1975)
\bibitem{fmrs} J. Feldman, J. Magnen, V. Rivasseau, R. Seneor, {\em Commun. Math. Phys.} {\bf 103}, 67-105 (1986)
\bibitem{fr} I. Frenkel, {\em J. Func. Anal.} {\bf 44,} 259-327 (1981)
\bibitem{froh}J. Frohlich.  {\em Phys. Rev. Lett.} {\bf 34,} 833-83 (1975)\\
--J. Frohlich, E. Seiler.  {\em Helv. Phys. Act.} {\bf 49,} 889-924 (1976)
\bibitem{gk} K. Gawedzki, A. Kupiainen, {\em Commun. Math. Phys.} {\bf 102}, 1-30 (1985)
\bibitem{gj} J. Glimm, A. Jaffe.  Quantum Physics.  Springer, 1987

\bibitem{salmhofer} M. Salmhofer.  Renormalization.  Springer Verlag, 1999.
\bibitem{cspaper} J. Weitsman, Fermionization and Convergent Perturbation Expansions in Chern-Simons Gauge Theory.  Preprint arxiv:0902.0097
\end{thebibliography}
\end{document}